\documentclass[aps,footinbib,superscriptaddress,twocolumn,prfluids]{revtex4-2}
\usepackage{amsmath,amsfonts, setspace,graphicx,comment,chemformula,bm}
\usepackage{soul,xcolor,algpseudocode,algorithm,amsthm}
\usepackage{multirow,hhline}
\usepackage[section]{placeins}

\newcommand{\be}{\begin{equation}}
\newcommand{\ee}{\end{equation}}

\newtheorem*{corollary*}{Corollary}

\def\RH#1{}
\usepackage{verbatim}
\usepackage{hyperref}

\begin{document}
\title{Optimal trajectories for Bayesian olfactory search in turbulent flows: the low information limit and beyond}
\author{R.\ A.\ Heinonen}
\affiliation{Dept.\ Physics and INFN, University of Rome ``Tor Vergata'', Via della Ricerca Scientifica 1, 00133 Rome, Italy}
\author{L.\ Biferale}
\affiliation{Dept.\ Physics and INFN, University of Rome ``Tor Vergata'', Via della Ricerca Scientifica 1, 00133 Rome, Italy}
\author{A.\ Celani}
\affiliation{Quantitative Life Sciences, The Abdus Salam International Centre for Theoretical Physics, 34151 Trieste, Italy}
\author{M. Vergassola}
\affiliation{Laboratoire de physique, \'Ecole Normale Sup\'erieure, CNRS, PSL Research University, Sorbonne University, Paris 75005, France}
\begin{abstract}
In turbulent flows, tracking the source of a passive scalar cue requires exploiting the limited information that can be gleaned from rare, stochastic encounters with the cue. When crafting a search policy, the most challenging and important decision is what to do in the absence of an encounter. In this work, we perform high-fidelity direct numerical simulations of a turbulent flow with a stationary source of tracer particles, and obtain quasi-optimal policies (in the sense of minimal average search time) with respect to the empirical encounter statistics. We study the trajectories under such policies and compare the results to those of the infotaxis heuristic. In the presence of a strong mean wind, the optimal motion in the absence of an encounter is zigzagging (akin to the well-known insect behavior ``casting'') followed by a return to the starting location. The zigzag motion generates characteristic $t^{1/2}$ scaling of the rms displacement envelope. By passing to the limit where the probability of detection vanishes, we connect these results to the classical linear search problem and derive an estimate of the tail of the arrival time pdf as a stretched exponential, which agrees with Monte Carlo results. We also discuss what happens as the wind speed decreases.
\end{abstract}

\maketitle
{\sc Introduction.} A wide variety of animals, from plankton \cite{roozen2001} to fruit flies \cite{alvarez2018} to dogs \cite{thesen1993}, combine olfactory sensing of odors with complex search behaviors which enable them to track the sources of the odors \cite{reddy2022}. A famous, striking example is the ability of moths to locate, starting from $\gtrsim 100$ m away, potential mates using a single sex pheromone cue to which they are sensitive via olfactory organs \cite{carde1979,kennedy1981,elkinton1987}. In many settings, including those relevant to flying insects \cite{murlis1992}, this olfactory search behavior occurs in a fully turbulent flow. Turbulent conditions are also relevant to chemical-sensing robots, which may be used to locate hazardous substances, fires or explosive devices \cite{martinez2006,martinez2013,nguyen2015,wiedemann2019}.

\begin{figure}
    \centering
    \includegraphics[width=\linewidth]{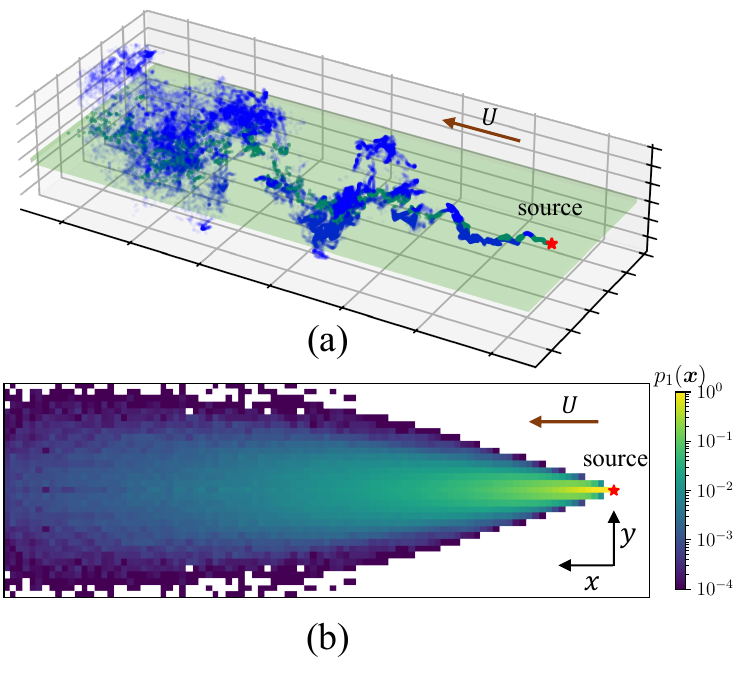}
    \caption{Upper panel (a): snapshot from the DNS showing particles (blue) and the 2--D slab (green) on which the data are coarse-grained. Lower panel (b): likelihood $p_1(\bm{x})$ of an encounter obtained from the DNS for $U/u_{\rm rms} \simeq 7.4.$}
    \label{fig:likelihoods}
\end{figure}

The consequences of turbulence in the olfactory search problem, by now well-studied \cite{yee1993,balkovsky2002,celani2014,reddy2022}, cannot be overstated. Odors may be modeled as passive scalars; the turbulent transport of passive scalars exhibits a remarkable degree of spatiotemporal intermittency, and the concentration fluctuations exhibit strongly non-Gaussian statistics \cite{shraiman_review}. As odors are emitted by a stationary and continuous source and mix into the surrounding flow, they form a characteristic concentration profile called a plume, whose shape depends on details of the flow. Within the plume, the concentration fluctuates wildly on account of the turbulence, forming a complex and patchy landscape (see Fig.~\ref{fig:likelihoods} for a visualization from simulation). If a searcher is sensitive to an odor above some detection threshold, its encounters with the odor are randomized and typically rare (except when the searcher is close to the source) \cite{celani2014}. Moreover, the odor typically manifests in isolated patches rather than a continuous field, rendering simple gradient-based search strategies effectively useless.

 How should a searcher\RH{, be it an animal or robot,} respond to these limited, random encounters in order to locate the source in minimal average time? Looking to biology can provide some inspiration. During upwind flight, a diverse array of flying insects exhibit a behavior called ``casting,'' which consists of broad zigzags perpendicular to the wind direction. Wind tunnel experiments suggest that this behavior is a very generic response to loss of contact with the odor cue \cite{kennedy1983,budick2006}. [In biological literature, ``zigzagging'' often refers to a specific insect behavior consisting of small-scale crosswind motion combined with upwind flight. We will frequently use the word ``zigzagging'' as simply a qualitative description of trajectories, without intending reference to this behavior.]

For robotics applications, this observation may be used to formulate biomimetic strategies, as in \cite{balkovsky2002}. Alternatively, if the searcher is equipped with a model for the spatially-dependent likelihood of an encounter, the search problem may be formalized as a Bayesian partially-observable Markov decision process (POMDP) \cite{kaelbling1998}. This formulation allows the \textit{optimal}, i.e., minimum average time, search strategy to be described by a dynamic programming (Bellman) equation, which may be approximately solved \cite{loisy2022,heinonen2023,loisy2023}.

The model likelihood is the key physics input into the problem. Previous work on the POMDP approach imposed a model likelihood \emph{ad hoc}, with the effects of turbulence captured by an effective diffusivity only. The realism of such a model is highly dubious in light of the complexity of turbulent transport; it is certain to fail, for example, in the large wind regime where ballistic transport dominates. (In the absence of a good model, the search problem could instead be attacked by model-free reinforcement learning techniques, as in Refs.~\cite{singh2023,rando2024}.)

In order to maximize the realism of the model, in this work we study optimal trajectories with respect to likelihoods taken directly from simulation data. In particular, we examine the trajectories which result from the absence of an encounter; because encounters are rare, in a typical search, most of the search time will be spent in this low-information phase.

The optimal trajectories which we compute are complex and depend strongly on the strength of the mean wind. We focus on the case where the mean wind is strong: in this case the optimal motion is to zigzag upwind similarly to casting. This motion exhibits characteristic diffusive scaling in both directions. At some critical time, if the searcher has still not found the source, it should return to its starting location. We explain these results using a simple model that assumes encounters are very rare except close to the source. The diffusive scaling is explained by a reduction to the classical linear search problem and is used to predict the arrival time pdf.

{\sc Methods.}   
From the POMDP point of view, the searcher is a Bayesian agent living on a gridworld which maintains a posterior probability density (or ``belief'') $b(\bm{x})$ over the possible relative positions of the source with respect to the agent; critically, this relative position is unknown. The searcher makes an observation at every timestep, which we define as measuring if the local concentration is above or below some threshold $c_{\rm thr}$. [In accordance with usual POMDP practice, the observation rate $\nu_\Omega$ is the same as the action rate, i.e.\ $\nu_\Omega= v/\Delta x,$ where $v$ is the speed of the searcher and $\Delta x$ is the grid spacing. In all figures, we will express time in units of $\nu_\Omega^{-1}$ and distance in units of $\Delta x.$] The belief is updated in two ways. When the searcher moves (``takes an action'' in POMDP jargon), the belief is transported according to
$ b(\bm{x}) \to b(\bm{x}-\delta \bm{x}),$
where $\delta \bm{x}$ is the searcher's displacement under the action. Secondly, when the searcher makes an observation $\Omega=\theta(c(\bm{x}) - c_{\rm thr}),$ it updates its belief according to Bayes' theorem \cite{box2011}:
\be
\label{eq:bayes}
b(\bm{x}) \to b(\bm{x}) p_\Omega(\bm{x}) /Z,
\ee
where $Z$ is the appropriate normalization constant and we have introduced the notation $p_\Omega(\bm{x}) \equiv {\rm Pr}(\Omega|\bm{x}).$
The likelihood $p_\Omega(\bm{x})$ must therefore be assumed to be known to the searcher. In the case where the mean wind $U\hat{x}$ is strong compared to flow velocity fluctuations, it can be modelled as proposed in \cite{celani2014}. The Lagrangian particle motion is predominantly ballistic on the short times relevant to this limit, which facilitates analysis of the tail of the concentration distribution. In particular, the probability that the instantaneous concentration is higher than $c_{\rm thr}$ is 
\be
\label{eq:celani}
p_1(\bm x) \propto \theta(x) x^{-\alpha} \exp\left[-\left(\frac{Uy}{u_{\rm rms} x}\right)^2  \right] \exp(-x/x_D), 
\ee
where $u_{\rm rms}$ is the typical fluctuation speed of the flow, $\theta$ is a Heaviside function, we have supposed the source is in $x=0$, $y$ is the distance from the plume centerline, $\alpha\ge0$ is a free parameter that depends on the environment, and $x_D$ is a characteristic decay length which depends on $c_{\rm thr}$. This result is only valid sufficiently far from the source. 

On the other hand, when the mean wind is small, long diffusive timescales become relevant to the Lagrangian particle motion and one instead expects
\be \label{eq:diffusive}
p_1(\bm{x}) \propto |\bm{x}|^{-1} \exp\left(-|\bm{x}|/\lambda \right) \exp( U x/2D),
\ee
with $D$ the effective (Taylor) diffusivity, $\lambda = \sqrt{D \tau/(1+U^2\tau/4D)},$ and $\tau = \tau(c_{\rm thr})$ is a typical time for a fluid puff, initially at the source, to grow large enough that the concentration is less than $c_{\rm thr}.$

Rather than impose one of these models directly, we used empirical models taken from simulations. We performed high-fidelity direct numerical simulations (DNS) of a turbulent flow (${\rm Re}_\lambda \simeq 150$), while tracking the positions of tracer particles which are emitted by a stationary point source. The system is advected by a uniform mean flow $U \hat{x};$ we repeated the simulation for $U/u_{\rm rms} \in \{0,1.2,2.5,4.9,7.4\}$. The tracers are taken as proxies for a passive scalar cue; this is a good approximation at high P\'eclet number. After coarse-graining the particle data onto a 2--D grid of about $\simeq 3250$ cells, we imposed a threshold concentration $c_{\rm thr}=200$ particles/cell and computed the likelihood $p_{1}(\bm{x})$ that this threshold is instantaneously exceeded (see Fig.~\ref{fig:likelihoods}). For the largest values of $U,$ the data are a good fit to Eq.~\ref{eq:celani} with $\alpha=1,$ and for the smallest values of $U$ the data fit Eq.~\ref{eq:diffusive} (see Fig.~3 of the Supplemental Material \footnote{See Supplemental Material at [hyperlink], which includes Refs.~\cite{shani_review,perseus}, for detailed information about methods for the DNS and POMDP solution, as well as additional calculations, plots, and movies.}).

For each flow, we then used the SARSOP algorithm \cite{sarsop} to extract a quasi-optimal search policy from the underlying dynamic programming equation \cite{sondik1978,kaelbling1998} (Supplemental Material, Eq. 3 \cite{Note1}). To study trajectories, we generated large ensembles of trajectories via Monte Carlo trials; for comparison, we did the same for the popular heuristic policy ``infotaxis'' \cite{infotaxis}, which seeks to minimize the entropy of the posterior. \RH{During search, observations were randomly drawn from the empirical likelihood. This artificially suppressed correlations between observations. It must be noticed that the no-hit trajectories under a policy that is optimized with respect to the uncorrelated statistics, as done here, will be entirely unchanged by the presence of correlations in the data during the search: only the arrival time statistics can change, since the optimal no-hit trajectory is a function of the likelihood and prior only. What happens when optimizing the policy including awareness of correlations is a technically different problem that will be discussed in a forthcoming paper.}

\begin{figure}
    \centering
    \includegraphics[width=\linewidth]{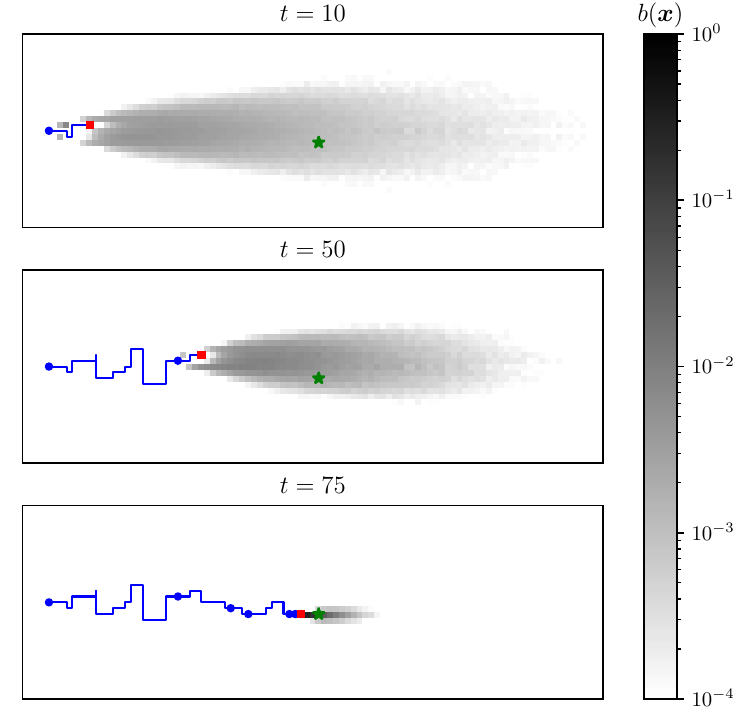}
    \caption{Snapshots from a trajectory produced from a quasi-optimal policy. The source is shown as a green star and the searcher as a red square. Each encounter is shown as a blue circle. The searcher's belief $b(\bm{x})$ on the position of the source is shown in grayscale. The searcher found the source a couple timesteps after $t=75$.}
    \label{fig:traj}
\end{figure}

An important detail is that all trajectories begin with an encounter at time zero, modeling the fact that the searcher has no incentive to search in the absence of an encounter. This induces a prior $b_0(\bm{x}) \propto p_1(\bm{x}).$ To ensure consistency, we begin each trajectory by randomly selecting a source position from the prior. We also constrain the searcher to move in two dimensions, which both eases the computational burden of the POMDP and simplifies the analysis of the results. An example quasi-optimal trajectory is shown in Fig.~\ref{fig:traj}.\\

{\sc Results.} Our results can be organized and understood through the following principle: because encounters are rare, the most important optimization the searcher must perform is to decide how it should move when it does not smell anything, and the single most likely trajectory is the one generated in the absence of any encounters after time zero. We will call this deterministic trajectory the ``no-hit trajectory.'' To support this claim, we evaluated the probability $P_{\rm nh}(t)$ that, at time $t,$ the searcher has not encountered the cue since time 0, given that it has not yet found the source. This is plotted for the largest wind speed in the upper panel of Fig.~\ref{fig:msd}; the results for other wind speeds are very similar and shown in Figs. 4--5 in the Supplemental Material \cite{Note1}. For both quasi-optimal and infotactic motion, $P_{\rm nh}(t) \gtrsim 0.5$ for all times $t$, so the no-hit trajectory plays an outsize role.

\begin{figure}
    \centering
    \includegraphics[width=\linewidth]{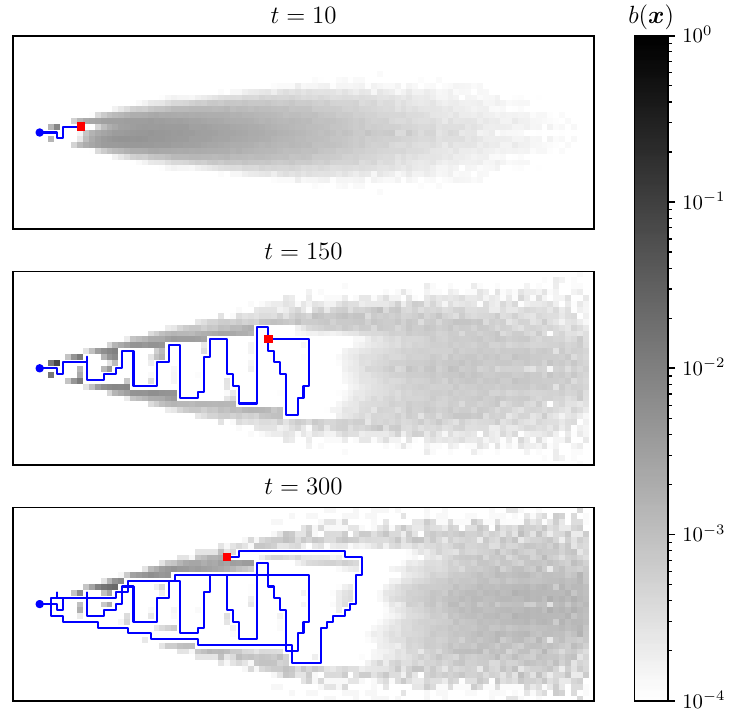}
    \caption{Quasi-optimal no-hit trajectory, for $U/u_{\rm rms}\simeq7.4$ The searcher is shown as a red square, the searcher's belief is shown in grayscale, and the starting point is a blue circle.} 
    \label{fig:zero_hit_traj}
\end{figure}

\begin{figure}
    \centering
    \includegraphics[width=\linewidth]{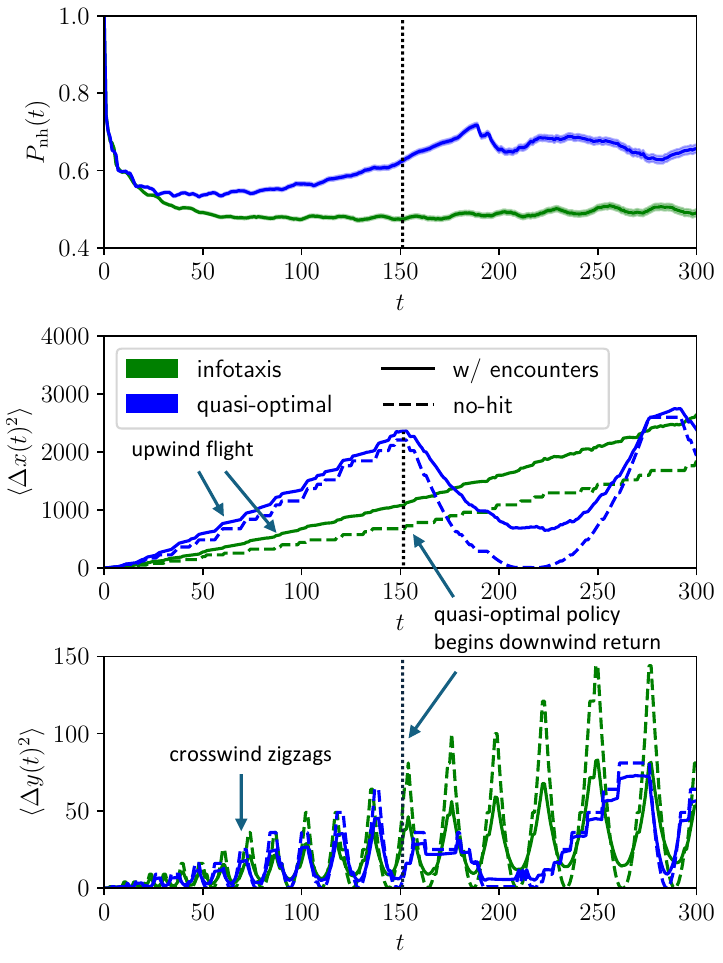}
    \caption{\RH{Search trajectory statistics for $U/u_{\rm rms} \simeq 7.4.$ Top panel: probability $P_{\rm nh}(t)$ that, at time $t,$ the searcher has not encountered the cue since time 0, given that it has not yet found the source; error bars are shown as the standard error. Lower panels: mean squared displacement (MSD) in the $x$ and $y$ directions.}} 
    \label{fig:msd}
\end{figure}

We find that when the mean wind is strong, the quasi-optimal no-hit trajectory begins by moving slowly upwind while zigzagging in the crosswind direction (Fig.~\ref{fig:zero_hit_traj}). This motion strongly resembles casting. However, at some critical time, the searcher ceases this behavior and returns downwind to the vicinity of its starting position. From then on, the motion is dominated by upwind and downwind motion. Downwind returns have also been occasionally observed in wind-tunnel experiments after a long time without an encounter \cite{willis1991,kuenen1994}, but this behavior is far less well-known. In contrast, infotaxis produces almost pure zigzagging motion in the absence of an encounter, and essentially never executes a downwind return.

Other examples of no-hit trajectories are supplied in Secs.\ VI C--D of the Supplemental Material \cite{Note1}. As the wind speed decreases, the downwind return occurs earlier and earlier, and zigzagging quickly becomes irrelevant to the optimal motion. Instead, at moderate wind speed, a quasi-optimal searcher exhibits repeated up- and downwind motion that begins to resemble a complex, asymmetric spiral. We suggest that the curves tend to approximately lie on isolines of the prior, which would imply the trajectories are essentially performing gradient ascent of the likelihood profile. When $U$ and $u_{\rm rms}$ are comparable, the problem has very little symmetry and, accordingly, the trajectories are especially complex. At $U=0,$ both the quasi-optimal and infotaxis trajectories resemble Archimedean spirals, consistent with previous studies \cite{infotaxis,barbieri2011,loisy2022}. 

These behaviors generate clear signatures in the mean squared displacement (MSD) of the searcher, $\langle \Delta x(t)^2\rangle \equiv \langle (x(t)-x(0))^2 \rangle$ and $\langle \Delta y(t)^2 \rangle \equiv \langle (y(t)-y(0))^2 \rangle,$ of the searcher, which we plot in Fig.~\ref{fig:msd} (lower panels). These are averaged over an ensemble of $10^4$ trajectories, and compared with the squared displacements resulting from a no-hit trajectory. The results are very similar, suggesting that motion in absence of an encounter is dominating the MSD signal. For the larger values of $U,$ the MSD in $y$ exhibits fast oscillations, the amplitude of which grows like $t$. The MSD in $x$ also grows approximately linearly. These results imply that the rms displacement envelopes in both $x$ and $y$ exhibit diffusive scaling: $x,y \sim t^{1/2}$. 

In the same figure, we also show the MSD resulting from an infotactic trajectory. While both infotaxis and quasi-optimal policy show the same behavior at small times, there is a clear qualitative difference at larger times: for the quasi-optimal policy, at some critical time, the oscillations and linear scaling cease, and from $t\simeq 150$--$200$ the MSD in $x$ decays rapidly like $t^2$, consistent with a ballistic downwind return. The time elapsed before returning becomes smaller as $U$ decreases (see Supplemental Material, Secs. VI A, C, D \cite{Note1}), and eventually the zigzag oscillations disappear completely. The infotactic policy, on the other hand, always produces an almost pure zigzag motion, except at the smallest nonzero value of $U$.\\

{\sc Low-information limit}. The results at large wind speed can be understood by means of a simplified model. In particular, we again assume the searcher initializes its belief as the prior $b_0(\bm{x})$ induced by the encounter at time zero, i.e., proportional to Eq.~\ref{eq:celani} for large enough $x.$ However, consistent with the dominance of the no-hit trajectory, we assume successive encounters only occur when the searcher is close to the source---the likelihood is very close to unity in a strip directly downwind of the source (see yellow region in Fig.~\ref{fig:likelihoods}, lower panel), which is where $\langle c \rangle > c_{\rm thr}.$ Empirically, this strip has negligible length except when $U$ is large; \RH{in the Supplemental Material (Sec.\ IV) we show that $\ell^* \propto U^{2/3}$ for large $U$ \cite{Note1}.}

Explicitly, we model the likelihood of an encounter after time $t=0$ as
\be \label{eq:low-info} p_1(\bm{x}) = \begin{cases} 1, & 0<x<\ell^*,\,y=0 \\ 0, & \textrm{otherwise,} \end{cases}\ee
where $x$ and $y$ are, respectively the upwind and crosswind position of the source with respect to the searcher, and $\ell^*$ is the characteristic length of the strip. The searcher uses this likelihood to update its belief. We will call this simplified model the \emph{low-information limit}. A similar model was used to describe sector search in Ref.~\cite{reddy2022sector}. 

\begin{figure}
    \centering
    \includegraphics[width=\linewidth]{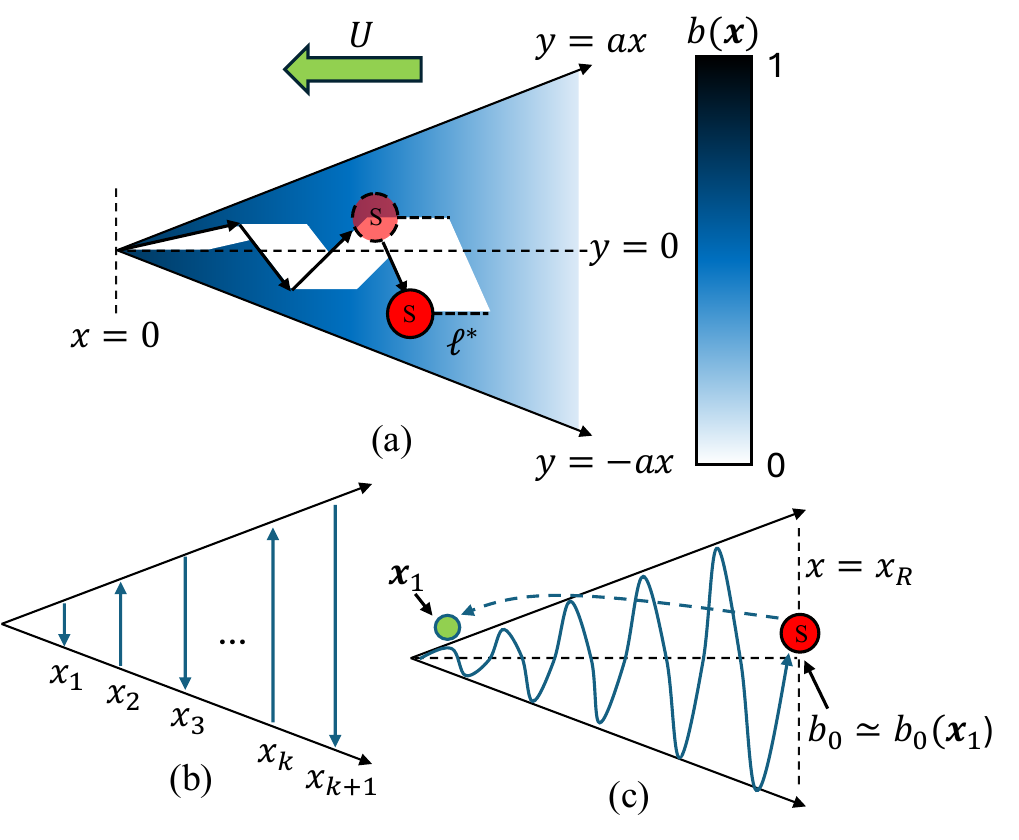}
    \caption{Upper panel (a): in the simplified model, the prior decays as a function of $x$ and is only supported on a cone symmetric about the $x$-axis. The searcher (red circle, labeled ``S'') sweeps out probability mass in the upwind direction as it moves, over a characteristic length $\ell^*.$ Lower left panel (b): a zigzag trajectory may be approximated by a sequence of crosswind motions at upwind distances $x_k,$ and the decision problem reduces to selecting the $x_k.$ Lower right panel (c): when the searcher has swept out the cone up to an upwind distance $x_R\sim x_D,$ the local value of the prior $b_0$ is comparable to that near the starting point and just outside the cone, $\bm{x}_1$. The searcher should return to $\bm{x}_1.$}
    \label{fig:lowinfo}
\end{figure}

In this limit, the result of the Bayesian update (Eq.~\ref{eq:bayes}) is that the searcher will simply zero out its prior in a $\ell^*$-neighborhood (and renormalize) as it moves (see Fig.~\ref{fig:lowinfo}, panel (a)). The optimal trajectory will then be the one that most efficiently explores the prior's support, so the geometry of the prior plays an essential role.

We now determine the optimal trajectories in the low-information limit. Due to the Gaussian factor in Eq.~\ref{eq:celani}, the prior will be strongly confined to a cone $C$ given by $|y| < a x$ with $a \sim u_{\rm rms}/U,$ so it is useful to approximate the confinement as exact: suppose the prior depends only on $x$ inside the cone, and it is zero outside (see Fig.~\ref{fig:lowinfo}, panel (a)).  We seek to minimize the expected time of arrival
\[ 
J = \frac{1}{2a} \int_{-a} ^{a} du \,\int_0^{\infty} dx \, p(x) T(x,u),\]
where $u\equiv y/x,$ $T(\cdot)$ is the time that the searcher first sweeps out a given point in space, and $p(x) \equiv x b_0(x).$

As a first observation, if $p(x)$ decays monotonically, then the optimal trajectory should (on average) visit states in order of increasing $x$ \footnote{To see this, note that if $x_2>x_1$ but $\int_{-a}^a du \, T(x_1,u) > \int_{-a}^a du \, T(x_2,u),$ a new assignment of arrival times $\tilde T$ with $\tilde T(x_1) = T(x_2), \tilde T(x_2) = T(x_1),$ and $\tilde T(x) = T(x)$ otherwise will result in a smaller value of $J$}. Second, the searcher should make as few turning points in $y$ as possible, since no probability mass is swept out if $dy/dt=0.$ Therefore it is reasonable to assume the only turning points occur at the edges of the cone. Such a trajectory will necessarily look like an upwind zigzag, but we have not yet determined the scalings of $x$ and $y$ with time.

A zigzag trajectory can be (for large enough $x$) modeled by a sequence of purely crosswind moves at positions $x_1,x_2,\dots$ (see Fig.~\ref{fig:lowinfo}, panel (b)). A key observation is that the searcher should not move further upwind than $\ell^*$ during any crosswind traversal, i.e., we have the constraint $x_{k+1}-x_k\le \ell^*$ for each $k$. Otherwise, there will be points not swept out by the trajectory, and because the trajectory is monotonic in $x,$ $J$ will be infinite.

In the absence of the inequality constraint, the decision problem of choosing the optimal sequence $\{x_k\}$ is exactly equivalent to the linear search problem (LSP), a well-studied decision problem which asks for the turning points of a trajectory on a line which minimizes the expected arrival time to a target drawn from some known probability density \cite{beck1964,beck1984,beck1986,alpern2003}.  This is shown in the Supplemental Material (Sec.\ V), after which we prove that if $p(x)$ decays exponentially or slower---as it does in our case---then $x_{k+1}-x_k$ asymptotically diverges for large $k$ \cite{Note1}. Thus the inequality constraint must saturate far enough upwind, i.e., $x_{k+1}-x_k = \ell^*,$ and one can show this gives rise to diffusive scaling as observed in the MSD data.

\RH{One may parametrize the resulting trajectory by a characteristic zigzag frequency $\omega$ and a distance traveled per half-period $\lambda.$ Note that the envelope of the trajectory defines a cone of exploration with half opening angle $\varphi=\arctan(2v/\lambda\omega),$ (here, $v$ is again the speed of the searcher). The preceding arguments suggest that, for a quasi-optimal policy, $\lambda \sim \ell^*$ and the opening angle should correspond to that of the belief cone, i.e., $\tan \varphi \sim u_{\rm rms}/U.$ This, in turn, implies $\omega \sim vU /\ell^* u_{\rm rms},$ leading to a model for the trajectory of the form} (ignoring numerical factors)
\begin{align}
\label{eq:zigzag}
x(t) \sim \sqrt{\frac{\ell^* v U t}{u_{\rm rms}}}; \; y(t) \sim \sqrt{\frac{\ell^* v u_{\rm rms} t}{U}} \cos \sqrt{\frac{v U t}{\ell^* u_{\rm rms}}},
\end{align}
with asymptotically constant speed $\sim v$ in an rms sense. \RH{Empirically, the quasi-optimal trajectory selects larger $\lambda$ but approximately equal $\omega$ as compared to infotaxis; this is a greedier choice that means the no-hit trajectory explores a narrower cone before the downwind return begins (the cones of exploration are compared in the Supplemental Material, Figs.\ 9--10 \cite{Note1}). }

We therefore conclude that the zigzag motion with $t^{1/2}$ scaling is a consequence of the fact that (a) the likelihood decays exponentially with $x$ (and not faster), (b) the likelihood decays faster than exponential outside the cone, and (c) the likelihood is small outside of a strip in front of the source. However, the argument (in particular, point (b)) rests on the approximation $\exp[-(u_{\rm rms} y/U x)^2] \ll \exp(-x/x_D).$ When the searcher reaches an upwind distance $x_R \sim x_D,$ this approximation breaks down, and the local prior probability will be comparable to that just outside the cone, near the starting point (see Fig.~\ref{fig:lowinfo}, panel (c)). This means that there is no continued benefit to search upwind, and the searcher should instead return downwind---as we indeed observe in the quasi-optimal policy.

We cannot analytically predict the optimal trajectories at finite $U$ (and moreover the approximation Eq.~\ref{eq:low-info} becomes untenable), but the low-information limit still grants some insights. Using the analysis of Ref.~\cite{celani2014}, one expects $x_D \sim u_{\rm rms}^{-1} (SU/c_{\rm thr})^{1/2}$ (where $S$ is the emission rate) when $U$ is large enough that ballistic single- and two-particle dispersion scaling still holds on relevant timescales. This is in basic agreement with our observation that downwind returns occur earlier as $U$ decreases. Eventually, as $U$ becomes comparable to $u_{\rm rms},$ diffusive motion begins to dominate trajectories of Lagrangian particles on timescales of interest. In this regime, the likelihood is no longer well-confined to a cone and gains some support even upwind of the source (this is visible in our simulations at $U/u_{\rm rms} \simeq 1.2$). This complexification of the prior geometry leads to concomitantly complex optimal trajectories (see Supplemental Material, Sec.\ VI A, C, D \cite{Note1}). 

Finally, a useful consequence of the $t^{1/2}$ scaling at large $U$ is that it allows us to predict the tail of the arrival time pdf analytically. In the low-information limit, the arrival time is dominated by the time of the first encounter after $t=0,$ which is the time it takes to reach a $\ell^*$-neighborhood of the source. Therefore, we can estimate the arrival time pdf $p(T)$ as
\be
p(T) = \int d^2 \bm{x} \, b_0(\bm{x}) \delta(T-T^*(\bm{x})),
\ee
where $T^*(\bm{x})$ is the time that the optimal trajectory first reaches $\bm{x}.$ $b_0(\bm{x})$ is proportional to the likelihood, which is given by Eq.~\ref{eq:celani} for large $U,$ and we have from Eq.~\ref{eq:zigzag} that $T^*(x,y) \sim (u_{\rm rms}/U)(x^2/v \ell^*).$ This gives
\be
p(T) \propto T^{-\alpha/2} \exp\left(-k \sqrt{T}\right),
\ee
with $k\sim x_D^{-1} \sqrt{U v\ell^*/u_{\rm rms}},$ a new prediction.

\begin{figure}
    \centering
    \includegraphics[width=\linewidth]{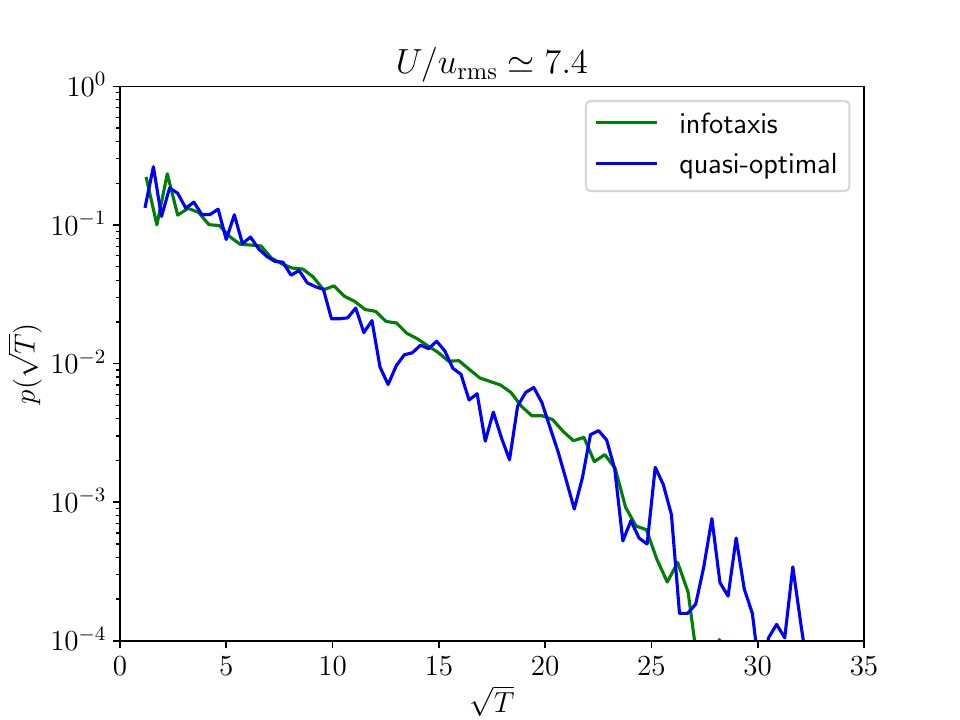}
    \caption{Pdf of the square root of the arrival time $\sqrt{T}$, for $U/u_{\rm rms} \simeq 7.4,$ obtained using $10^5$ Monte Carlo trials.}
    \label{fig:v9_pdf}
\end{figure}

In Fig.~\ref{fig:v9_pdf} we plot the pdf of $\sqrt{T}$ for $U/u_{\rm rms} \simeq 7.4$, where $T$ is the time of arrival. There is a broad range $0\le T \lesssim 500$ where the pdf appears to decay exponentially, implying $p(T) \sim \exp(-k \sqrt{T})$ as predicted. The agreement with exponential decay is more precise for infotaxis; we could have anticipated this since infotaxis was observed to obey $t^{1/2}$ scaling for the entire trajectory, whereas this scaling breaks down for the quasi-optimal policy once it begins its downwind return. 

\RH{\sc{Conclusion}.} In summary, our results suggest that the ubiquity of casting as a response to loss of plume contact may be due to its being the optimal no-hit trajectory in the limit of low information, in the presence of strong mean wind. The same limit makes nontrivial quantitative predictions about the temporal scaling of the trajectories and the arrival time pdfs, which agree with results obtained using quasi-optimal policies extracted from the dynamic programming equation.

\RH{Finally, while it is not the goal of this paper to precisely model the behavior of any specific animal, it is nevertheless useful to discuss the relationship between our results and insect experiments. To our knowledge, Ref.~\cite{kuenen1994} provides the most detailed quantitative experimental results characterizing moth behavior after plume loss during upwind olfactory search. These experiments found, for one, that crosswind excursion distances increase roughly linearly with the period number, which agrees with our results and simply reflects the conical shape of the plume in the strong wind regime. On the other hand, that study also found that the upwind distance traveled decreases monotonically with period, whereas in our study this was found to be constant under a quasi-optimal strategy. There are a number of possible explanations for this discrepancy --- it could be related to the fact that the experiments were performed in three dimensions whereas we were constrained to two, or the fact that our searchers did not have access to instantaneous anemometric cues, or perhaps moths simply do not implement a quasi-optimal Bayesian strategy.} 

\RH{There is essentially no quantitative data provided on downwind returns in either of Refs.~\cite{willis1991} and \cite{kuenen1994}, which are to our knowledge the only two works which specifically reported this behavior in the context of plume loss. In Ref.~\cite{kuenen1994}, the behavior is reported anecdotally; in Ref.~\cite{willis1991}, it is described as the consistent behavior of moths exhibited after a long time without contact with the plume. This latter result is qualitatively consistent with our findings on Bayesian searchers in an environment with large mean wind. Additional experiments would be necessary to establish quantitative agreement. It would be important for any such experiment to distinguish between downwind returns after plume loss from downwind flight executed in the service of anemometry (Refs.~\cite{guichard2010,stupski2024} discuss downwind flight in this context). One would expect anemometric downwind flight to occur much earlier in the trajectory and to not be correlated with plume loss.}

\begin{acknowledgments}
We thank Aurore Loisy, Massimo Cencini, Lorenzo Piro, and Fabio Bonaccorso for fruitful discussions and acknowledge useful interactions at the 2023 Les Houches summer school on turbulence. This work received funding from the European Union's Horizon 2020 Program under grant agreement No.\ 882340. We also acknowledge financial support under the National Recovery and Resilience Plan (NRRP), Mission 4, Component 2, Investment 1.1, Call for tender No. 104 published on 2.2.2022 by the Italian Ministry of University and Research (MUR), funded by the European Union – NextGenerationEU– Project Title Equations informed and data-driven approaches for collective optimal search in complex flows (CO-SEARCH), Contract 202249Z89M. – CUP B53D23003920006 and E53D23001610006. \RH{The data that support the findings of this article are openly available \cite{turbodor,code}, embargo periods may apply.}
\end{acknowledgments}
%

\end{document}


\title{Supplemental Material: Optimal trajectories for Bayesian olfactory search in turbulent flows: the low information limit and beyond}
\author{R.\ A.\ Heinonen}
\affiliation{Dept.\ Physics and INFN, University of Rome ``Tor Vergata'', Via della Ricerca Scientifica 1, 00133 Rome, Italy}
\author{L.\ Biferale}
\affiliation{Dept.\ Physics and INFN, University of Rome ``Tor Vergata'', Via della Ricerca Scientifica 1, 00133 Rome, Italy}
\author{A.\ Celani}
\affiliation{Quantitative Life Sciences, The Abdus Salam International Centre for Theoretical Physics, 34151 Trieste, Italy}
\author{M. Vergassola}
\affiliation{Laboratoire de physique, \'Ecole Normale Sup\'erieure, CNRS, PSL Research University, Sorbonne University, Paris 75005, France}

\maketitle
\section{Methods}\label{app:methods}
\subsection{DNS}\label{app:DNS}
In order to obtain realistic encounter likelihoods, we solved the incompressible, 3--D Navier-Stokes equations
\begin{align}
\partial_t \bm{u} + (\bm{u} \cdot \nabla) \bm{u} &= - \nabla p + \nu \nabla^2 \bm{u} + f, \\
\nabla \cdot \bm{u} &= 0,
\end{align}
under turbulent conditions with $\mathrm{Re}_\lambda \simeq 150.$ Here, $f$ is a random isotropic forcing at the smallest nonzero wavenumbers of the system, with a correlation time of 160 simulation timesteps, or approximately one Kolmogorov time $\tau_\eta$. Using a pseudospectral code dealiased according to the two-thirds rule, the system was solved on a $1024\times512\times512$ grid, with a uniform spacing $\delta x = \delta y = \delta z \simeq \eta$ (with $\eta$ the Kolmogorov scale), and periodic boundary conditions in all three directions. The timestepping was performed using the second-order explicit Adams-Bashforth method. The system was advected by a uniform mean wind $\bm{U} \approx -2.5 u_{\rm rms} \hat{x},$ where $u_{\rm rms}$ is the rms speed of the flow in the comoving frame of the wind, and $\hat{x}$ is the elongated axis of the grid. We produced the mean wind by means of a Galilean transformation.

Five stationary point sources each emitted 1000 Lagrangian tracer particles every 10 simulation timesteps, which corresponds to every $\approx 1/15 \tau_\eta$. The fluid velocities $\bm{u}$ at the particle positions were obtained using a sixth-order B-spline interpolation scheme and then used to evolve the particle positions $\bm{X}$ in time according to $\dot{\bm{X}}=\bm{u}(\bm{X},t)$ over an infinite lattice of copies of the periodic flow. Their positions, velocities, and accelerations were tracked and dumped every $\tau_\eta,$ for a total of about 3000 timesteps. Each source of particles was treated as independent, and we averaged our results over them for the purpose of achieving better statistics.

\begin{figure}
    \centering
    \includegraphics[width=0.5\linewidth]{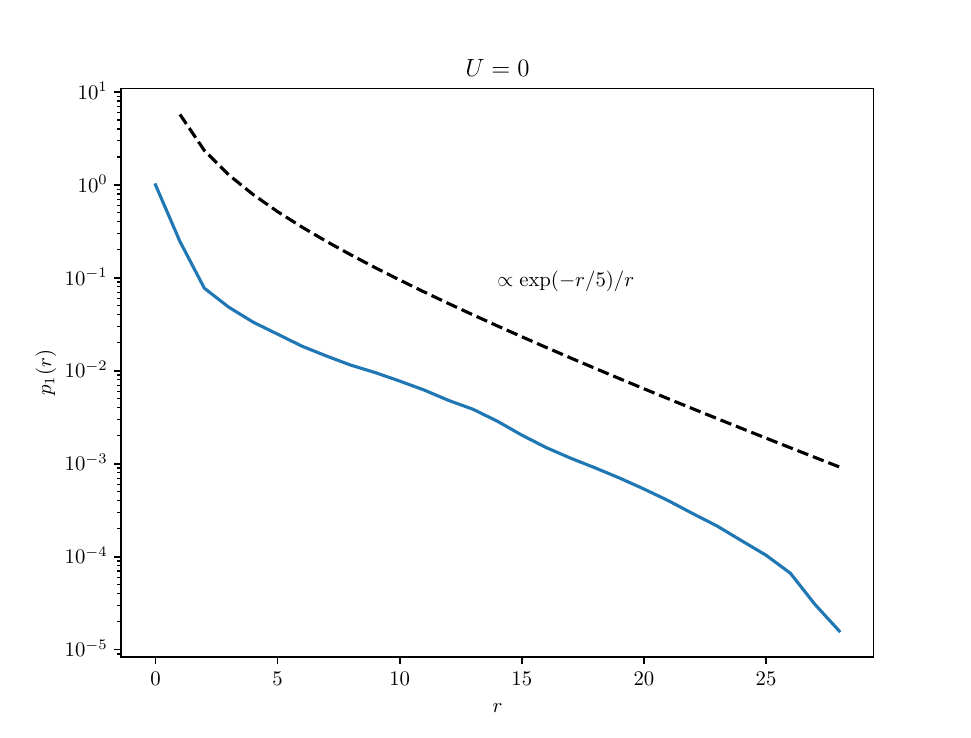}
    \caption{Symmetrized empirical likelihood $p_1(r)$ of encounter in isotropic setting. For comparison, we also show a diffusive model, which fits reasonably well.}
    \label{fig:isotropic_likelihood}
\end{figure}

\begin{figure}
    \centering
    \includegraphics[width=0.7\linewidth]{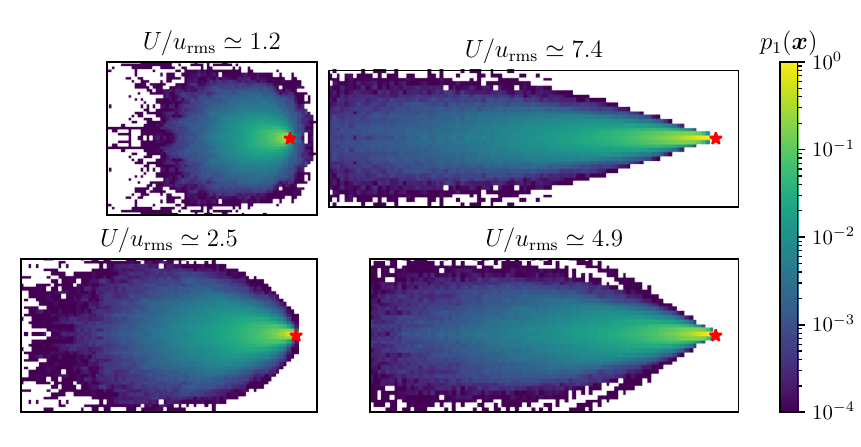}
    \caption{Empirical likelihoods $p_1(\bm{x})$ of encounter in the presence of a mean wind $U$.}
    \label{fig:windy_likelihoods}
\end{figure}

The particle data were then coarse-grained onto a quasi-two-dimensional slab which contained the source and was parallel to the wind. The slab comprised $\simeq 3250$ cubic cells with side length $\simeq 15 \eta$ (here, $\eta$ is the Kolmogorov length); the aspect ratio of the grid depended on the wind speed and was chosen to ensure the likelihood was $<10^{-4}$ at the boundary. The likelihoods were computed as the time averaged probability that the number of particles in a given cell exceeded threshold. In the presence of wind, these were symmetrized along the wind axis; in the isotropic case, they were symmetrized to be a function of $r$ only. We show likelihoods for all windspeeds in Figs.~\ref{fig:isotropic_likelihood}--\ref{fig:windy_likelihoods}.

\subsection{POMDP details}\label{app:POMDP}
\begin{table}[]
    \centering
    \begin{tabular}{|c|c|}
    \hline
        states & agent position $\bm{x}$ w.r.t.\ source in gridworld  \\
        \hline
        actions & move to an adjacent 
        gridpoint \\ \hline
        state transitions & deterministic, determined by actions \\ \hline
        observations & binary (above/below threshold) concentration measurements \\ \hline
        reward & unit penalty per timestep until source reached \\ \hline
        prior (initial belief) & induced by encounter at time zero \\ \hline
        
    \end{tabular}
    \caption{Summary of olfactory search POMDP.}
    \label{tab:pomdp}
\end{table}
The POMDP was setup similarly to Refs.~\cite{loisy2022,heinonen2023,loisy2023}, to which we refer the interested reader for details. The agent lived on the gridworld defined in the previous subsection, and the source was fixed at an unknown location. Its belief $b$ was a probability distribution over all possible relative positions between the source and agent; these positions are the states of the POMDP. At each timestep the agent drew a binary observation from the likelihood, which was used to update its belief via Bayes' theorem. The agent then took an action, defined as moving to one of the four adjacent grid cells (at the boundary, actions which would take the agent out of the gridworld were excluded). The motion of the agent was treated as deterministic (i.e.\ the agent successfully moved to the intended grid point with probability one).

Monte Carlo trials were initialized by placing the agent at a chosen starting location, setting the prior by triggering an encounter at time zero, and drawing the relative position of the source from the prior for self-consistency. The trials terminated when the agent arrived at the source, or if the agent had still not arrived after 2500 timesteps.

The reward function was modeled as a unit penalty per timestep until the source was reached. The state transitions, observation likelihood, reward function, and choice of prior (see Table~\ref{tab:pomdp} for a summary) were then used as inputs into the SARSOP algorithm \cite{sarsop}, which approximately solves the dynamic programming equation, also called the Bellman equation \cite{sondik1978,kaelbling1998}. This equation expresses the recursive structure of the ``value function'' $V(b)$, which returns the total expected reward conditioned on the current belief being $b$. For our problem it reads
\be \label{eq:bellman}
V(b) = -1 + \gamma \max_{\bm{a}\in A}  \sum_{\Omega\in\{0,1\}} {\rm Pr}(\Omega | b,\bm{a}) V(b'(\cdot | \Omega,\bm{a})), 
\ee
where $\gamma\in(0,1]$ is the discount factor, which geometrically suppresses future rewards in favor of immediate ones, $\Omega$ is an observation, $A$ is the set of four possible actions, $b'(\cdot | \Omega,\bm{a})$ refers to the belief obtained after taking action $\bm{a},$ observing $\Omega,$ and then applying a Bayesian update to $b,$ and
\begin{equation}
{\rm Pr}(\Omega |b,\bm{a}) = \sum_{\bm{x}} p_{\Omega}(\bm{x}+\bm{a}) b(\bm{x}),
\end{equation}
with $p_\Omega$ the likelihood as defined in the main text. Knowledge of $V(b)$ immediately yields the optimal policy $\pi^*(b)$, which is just the maximizing action on the RHS of Eq.~\ref{eq:bellman}. 

SARSOP obtains an approximation for the value function
\be
V(b) \simeq \max_k b \cdot \bm{\alpha}_k,
\ee
where the dot product indicates a product and sum over state indices, and $\{\bm{\alpha}_k\}$ is a collection of matrices with the same dimension as the state space. Each $\alpha_k$ comes equipped with an associated action $a_k,$ and the (quasi-)optimal policy selects the action associated with the maximizing $\bm{\alpha}_k.$ \RH{This is a piecewise-linear and convex approximation to the value function, which can be made arbitrarily accurate in principle \cite{sondik1978}. Like other ``point-based'' methods \cite{shani_review}, SARSOP extracts the $\bm{\alpha}_k$'s by performing a form of value iteration (called the ``backup'' operation) at specific points in belief space. In SARSOP, these points are found by building a tree of beliefs reachable from the prior by taking actions and making observations, and pruning from this tree beliefs which will not be reached by an optimal policy. This gives it an advantage over other point-based methods such as Perseus \cite{perseus}, which requires an ad hoc prescription for collecting belief points. We note in passing that a visual inspection of the $\bm{\alpha}_k,$ once obtained, did not reveal any immediately interpretable pattern.}

The SARSOP algorithm requires $\gamma<1$; we took $\gamma=0.98,$ as close to unity as possible without compromising the successful convergence of the algorithm.

\RH{SARSOP will continue to refine the policy indefinitely and must be terminated by the user. On this problem, we found that the empirical performance of the policy---whether measured by the reward being optimized by the algorithm or the mean arrival time---stops improving substantially beyond a walltime of around 2000 seconds. To give the reader an idea of this behavior, we present the performance as a function of walltime in Table \ref{tab:sarsop} for $U/u_{\rm rms}\simeq 7.4$; of course, these results will be machine-dependent. For the presented results, we chose to terminate the algorithm and outputted a policy after a walltime of 4000 seconds, which produced the best measured mean arrival time in the $U/u_{\rm rms}\simeq 7.4$ case. Note that stopping the algorithm at a different time may produce somewhat different quantitative results, but we found that the important qualitative features of the trajectories are robust.}

\begin{table}[]
    \centering
    \begin{tabular}{|c|c|c|}
    \hline
        walltime (s) & $\langle T \rangle $ & $\langle R \rangle$  \\
        \hline
         30 & $49.31 \pm 0.86$ & $19.67 \pm 0.16$ \\
         100 & $46.83 \pm 0.85$ & $19.32 \pm 0.16$ \\
         300 & $47.34 \pm 0.85$ & $19.58 \pm 0.16$\\
         1000 & $46.64 \pm 0.83$ & $19.28 \pm 0.16$\\
         2000 & $45.47 \pm 0.81 $ & $19.34 \pm 0.16$\\
         {\bf 4000} & $\bm{44.58 \pm 0.76}$ & $\bm{19.29 \pm 0.15}$ \\
         6000 & $ 44.86 \pm 0.79$  & $19.08 \pm 0.16$ \\
         8000 & $44.61 \pm 0.79$ & $19.15 \pm 0.15$\\
         10000 & $45.67 \pm 0.83$ & $19.23 \pm 0.15$ \\
         \hline
    \end{tabular}
    \caption{\RH{Table showing empirical SARSOP performance as function of walltime for $U/u_{\rm rms} \simeq 7.4$, which we measure by the mean arrival time $\langle T \rangle$ and the mean reward $\langle R \rangle,$ where $R(T)= (1-\gamma^T)/(1-\gamma)$ and $\gamma=0.98.$ The results were obtained from sets of $10^4$ Monte Carlo trials, and uncertainties are expressed as standard errors on the mean. Note that the empirical failure rate (i.e., the probability that $T\ge 2500$) was always zero. The policy we used to generate figures is shown in boldface.}}
    \label{tab:sarsop}
\end{table}

\section{Estimate of likelihood for small mean wind}\label{app:likelihood}
In the Lagrangian picture, the concentration due to a continuously-emitting point source at the origin with emission rate $S$ is \cite{shraiman_review}
\be
c(\bm{x}) = S \int_{0}^\infty dt \, {\rm Pr}(\bm{X}(t)=\bm{x}| \bm{X}(0)=0).
\ee
${\rm Pr}(\bm{X}(t)=\bm{x}| \bm{X}(0)=0)$ is the probability that a Lagrangian particle starting at the origin arrives at the test point $\bm{x}$ at time $t$ and is generally represented by a path integral. The integration variable $t$ then represents how long in the past the particle was released. In the large mean wind limit, one may exploit the ballistic scaling of the particles to only consider the contribution from times $t=x/U$, which greatly simplifies the analysis and leads to the results of Ref.~\cite{celani2014}. 

However, if the mean wind is small, all past times contribute to the concentration and the calculation is in principle quite involved. In particular, we are interested in the tail of the distribution of $c(\bm{x}).$

To estimate $p_1(\bm{x}) = {\rm Pr}(c>c_{\rm thr}|\bm{x})$ for large $c_{\rm thr},$ we assume large fluctuations in the concentration are due to a puff released from the source remaining small for a much longer time than usual. Fluid velocity fluctuations on the scale of the puff size will tend to disperse the puff, increasing its radius and thereby decreasing the local concentration. The probability that the puff remains small some time $t^* \gg \tau,$ where $\tau=\tau(c_{\rm thr})$ is the typical time for the concentration of the puff to be reduced below $c_{\rm thr},$ will be the product of many largely independent factors, leading to Poisson statistics and a tail of the form $p(t^*) \sim \exp(-t^*/\tau).$ This gives
\be
{\rm Pr}(c>c_{\rm thr}|\bm{x}) \simeq S \int_0^\infty dt \, \int_t^\infty dt^* p(t^*) {\rm Pr}(\bm{X}(t)=\bm{x}| \bm{X}(0)=0).
\ee

At sufficiently large $t$, the particle motion decorrelates and passes from ballistic to diffusive, leading to 
\[ {\rm Pr}(\bm{X}(t)=\bm{x}| \bm{X}(0)=0) = (4 \pi D t)^{-3/2} \exp(-|\bm{x-Ut}|^2/4Dt)\]
for a characteristic diffusivity $D.$ This leads to 
\be \label{eq:diffusive}
p_1(\bm{x}) \propto |\bm{x}|^{-1} \exp\left(-|\bm{x}|/\lambda \right) \exp( U x/2D),
\ee
with $\lambda = \sqrt{D \tau/(1+U^2\tau/4D)}.$ This is proportional to the solution of the advection diffusion equation
\[
\partial_t p_1 + U \partial_x p_1 = D \nabla^2 p_1 + S \delta(\bm{x}).
\]

\section{Collapse of empirical likelihoods to models}
\begin{figure}
    \centering
    \includegraphics[width=0.5\linewidth]{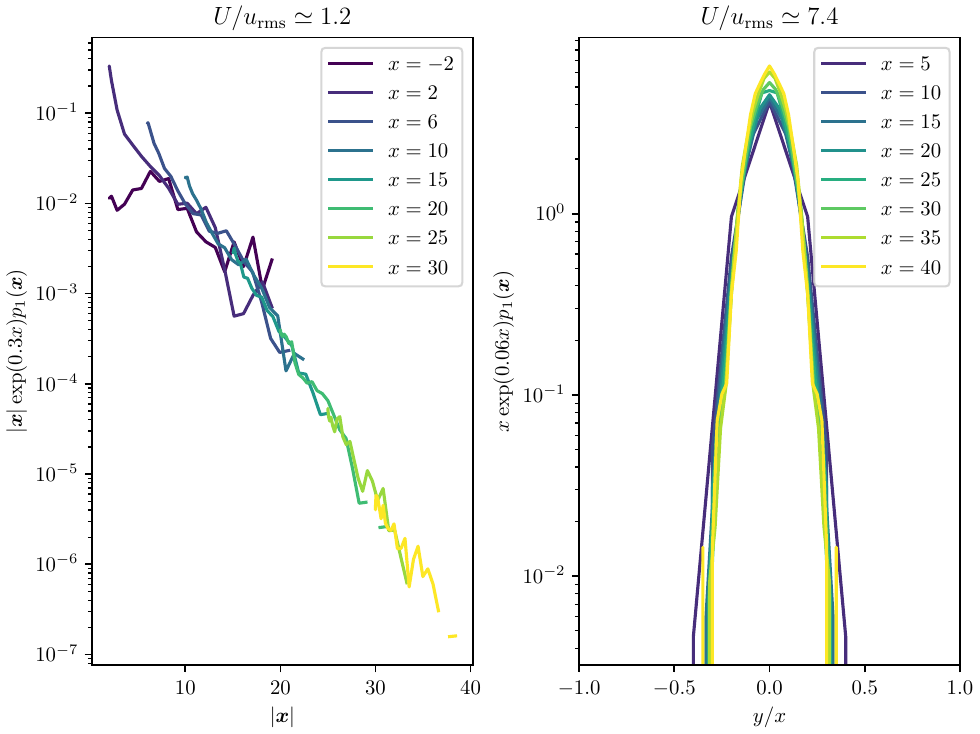}
    \caption{Left panel: collapse of likelihood to a model of the form Eq.~\ref{eq:diffusive}, for the smallest wind speed $U/u_{\rm rms} \simeq 1.2.$ Right panel: collapse of likelihood to a model of the form Eq.~\ref{eq:celani} with $\alpha=1$ for the largest wind speed $U/u_{\rm rms} \simeq 7.4.$  }
    \label{fig:collapse}
\end{figure}

In the left panel of Fig.~\ref{fig:collapse}, we show that the likelihood at the smallest wind speed $U/u_{\rm rms} \simeq 1.2$ is a good fit to a model of the form Eq.~\ref{eq:diffusive}, indicating that diffusive effects are dominant. On the other hand, the right panel shows the collapse of the likelihood for the largest wind speed $U/u_{\rm rms} \simeq 7.4$ onto a model of the form
\be \label{eq:celani}
p_1(\bm{x}) = x^{-\alpha}\exp(-x/x_0) \exp\left[-c\left(\frac{y}{x} \right)^2\right],
\ee
for $x>0$ (downwind), as predicted in Ref.~\cite{celani2014} and discussed in the main text. In particular, the data are a good fit to $\alpha=1.$

\RH{\section{Scaling of $\ell^*$}
 Let the source have size $\ell_S$ and emission rate $S.$ The width of the strip is $\sim \ell_S$ to leading order in $u_{\rm rms}/U$. We now estimate the strip's length, $\ell^*.$  If one estimates the typical concentration at downwind distance $x$ from the source, $\ell^*$ can be understood as the scale where this concentration crosses over with $c_{\rm thr}.$ Explicitly, for large mean wind, the concentration at the source scales as
 \[ c_S \sim \frac{S}{U \ell_S^2}, \]
 which comes from balancing the injection with the outward flux. After release from the source, a puff will grow as 
 \[ r \sim (k t)^{1/\gamma} \sim \left(\frac{k x}{U}\right)^{1/\gamma}\]
 for some exponent $\gamma>0$ and constant $k.$ This leads to an estimate of the typical concentration
 \[ c\sim \frac{c_S \ell_S^3}{r(x)^3},\]
so $c_{ \rm thr} \sim c_S \ell_S^3/r(\ell^*)^3$ by the definition of $\ell^*$ Rearranging gives 
\[ \ell^* \sim \frac{U}{k} \left(\frac{S \ell_S}{U c_{\rm thr}}\right)^{\gamma/3}.\]
For short times after release (i.e., large $U$) it is reasonable to assume ballistic (Batchelor) separation, i.e., $\gamma\simeq1$ and $k\sim u_{\rm rms},$ so we expect
\[ \ell^* \sim \frac{U^{2/3}}{u_{\rm rms}} \left(\frac{S \ell_S}{c_{\rm thr}} \right)^{1/3}.\]
Let us notice furthermore that $\ell^*,$ which is connected to the physics near the source, is not directly related to the likelihood decay length, $x_D$, which is instead connected to the statistics of rare events far downwind of the source.}

\section{Analysis of linear search problem}\label{app:math}
\subsection{Reduction to linear search} 
We first note that the conical search problem, treated in the main text as a model for the $U/u_{\rm rms}\to\infty$ limit, is topologically the same as the search problem on the isotropic (i.e., $U=0$) plane, with a cut along a semiaxis, under the identification $x\leftrightarrow r, u \leftrightarrow\theta, a \leftrightarrow\pi$ (where $r$ and $\theta$ are the polar coordinates in the plane). Zigzag trajectories with $t^{1/2}$ scaling in the former correspond to Archimedean spirals in the latter. We find the isotropic problem is a bit more convenient to study; our goal is to show that the trajectory is an Archimedean spiral.

Let the (isotropic) prior be $b_0(r)$. Put $p(r) = 2\pi r b_0(r)$ so that $\int dr \, p(r) = 1.$ Assume $p(r)$ decreases monotonically, has finite mean, and is nonzero everywhere. The polar coordinates of the source are $(r^*,\theta^*)$ drawn respectively from $b_0(r)$ and a uniform distribution on $[0,2\pi).$ The agent finds the source when its coordinates $(r,\theta)$ satisfy $\theta = \theta^*$ and $|r-r^*|<  \ell^*/2,$ where $\ell^*$ is the characteristic length swept out by the agent (see Fig. 6 in the main text). 

Analogizing the arguments of the main text, the agent should follow a trajectory monotonic in $r$ and with no turning points in $\theta$. This means the trajectory is a spiral and can be parametrized as $r=r(\theta).$ We also have the constraint
\be \label{eq:constraint}
r(\theta +2\pi) \le r(\theta) + \ell^*
\ee
or $\langle r'(\theta) \rangle \le \ell^*.$

When $r\gg \ell^*$ we have
\[
v^2 = \dot r^2 + r^2 \dot \theta^2 = \dot \theta^2 (r'(\theta)^2 + r^2) \simeq r^2 \dot \theta^2,
\]
so the motion is in the $\hat \theta$ direction, and the trajectory may be approximated by a sequence of concentric circles.

Asymptotically, the decision problem may then be approximately reduced to choosing a discrete sequence of radii $r_k$ and sweeping out each circle $r= r_k.$ The objective function for the $r_k$ is then
\begin{align}
J &= \sum_{k=1}^\infty \int_{ r_{k-1}+\ell^*/2}^{r_k+\ell^*/2} dr \, p(r) \left[\pi r_k + 2\pi \sum_{j=1}^{k-1} r_j \right] \nonumber \\ &= \sum_{k=1}^\infty \int_{ \tilde r_{k-1}}^{\tilde r_k} dr \, p(r) \left[\pi \tilde r_k + 2\pi \sum_{j=1}^{k-1} \tilde r_j \right] + \mathrm{const.} \nonumber \\ &= \pi \sum_{k=1}^\infty \tilde r_k ( G(\tilde r_k) + G(\tilde r_{k-1})) + \mathrm{const.}
\end{align}
where we have defined $r_0 = 0,$ $\tilde r_i = r_i + \ell^*/2,$ and $G(r) = \int_r^\infty dr' p(r').$ Note that the agent finds the source instantly if $r^*<\ell^*/2.$

In the absence of the constraint $\tilde r_{k+1}-\tilde r_k \le \ell^*,$ this problem has exactly the same objective (up to an irrelevant affine transformation) as the linear search problem (LSP) \cite{beck1964,beck1984,beck1986,alpern2003}. It is therefore useful to study solutions of the LSP. We have
\be 
\frac{1}{\pi} \frac{\partial J}{\partial\tilde r_
i}
= G(\tilde r_i) + G(\tilde r_{i-1}) - p(\tilde r_i)(\tilde r_i +\tilde r_{i+1}) 
\ee
where $G(r) = \int_r^\infty dr' p(r').$ Therefore, in the absence of the constraint, an optimal choice of the $\tilde r_k$ must satisfy $\partial J/\partial\tilde r_i=0,$ whence we have the recurrence
\be
\label{eq:recursion}
\tilde r_i + \tilde r_{i-1} =  \frac{G(\tilde r_i) + G(\tilde r_{i-1})}{p(\tilde r_i)}.
\ee

\subsection{Sufficient condition for saturation of Eq.~\ref{eq:constraint}} The critical question is, for what distributions $p(r)$ do the optimal $\tilde r_k$ satisfy
\[
\tilde r_{k+1} - \tilde r_k \ge \ell^*
\]
for sufficiently large $k$ in the unconstrained problem? This leads to $\tilde r_{k+1} - \tilde r_{k} = \ell^*$ upon reintroducing the constraint, which in turn leads to diffusive scaling (see below).

While we do not attempt to answer this question completely, it is possible to show that probability densities which decay exponentially or slower eventually satisfy this inequality. The sketch of the proof is as follows: first, it is straightforward to show that, for any sufficiently well-behaved probability density $p(r)$ on $(0,\infty)$ with finite mean and $p(r)>0$ everywhere, 
\be \label{eq:lhopital} p(r)^{-1} \int_r^\infty dt \, p(t) \le Cr,\ee
for some $C\in(0,1)$ and for sufficiently large $r$. This follows after application of L'H\^opital's rule to the expression $\int_r^\infty dt \, p(t)/rp(r).$ 

We now assume $p(r)$ decays exponentially or slower---that is, suppose that if $x>0,$
\be \label{eq:expdecay} p(r+x)/p(r) \ge \exp(-x/a)\ee
for large enough $r$. Then using the recursion Eq.~\ref{eq:recursion} and Eq.~\ref{eq:lhopital}, 
\begin{align*}
2 \tilde r_k &\le \tilde r_k + \tilde r_{k+1} \\ & = 2 p(\tilde r_k)^{-1} \int_{\tilde r_k}^\infty dr \, p(r) + \int_{\tilde r_{k-1}}^{\tilde r_k} dr \, \frac{p(r)}{p(\tilde r_k)} \\ & \le 2C \tilde r_k  + \int_{\tilde r_{k-1}}^{\tilde r_k} dr \, e^{(\tilde r_k - r)/a} \\ &=2C \tilde r_k + a\left(e^{(\tilde r_k - \tilde r_{k-1})/a} -1 \right).
\end{align*}

Rearranging, we have
\be
\tilde r_k - \tilde r_{k-1} \ge a \log\left(\frac{2(1-C)\tilde r_k}{a} +1 \right).
\ee
The RHS grows without bound, since the optimal $\tilde r_k$ are themselves increasing and unbounded. Therefore, in the absence of the constraint, $\tilde r_k - \tilde r_{k-1} \ge \ell^*$ for large enough $k.$

For our olfactory search problem, for large $U,$ the correct $p(r)$ to consider is $p(r) =x_0^{-1} \exp(-r/x_0)$ (see Eq.~\ref{eq:celani}). Equation~\ref{eq:expdecay} then holds trivially, and we can conclude that, indeed, $\tilde r_{k+1} -\tilde r_k = \ell^*$ asymptotically. The result also holds e.g.\ for power laws $p(r)\propto r^{-\alpha}$ for $\alpha>2.$

\subsection{Linear growth of the $\tilde r_k$ leads to diffusive scaling}
Having established that $\tilde r_{k+1}-\tilde r_k = \ell^*$ for large enough $k,$ it is easy to see that $t^{1/2}$ scaling follows inevitably. We have 
\[
\dot r = r'(\theta) \dot \theta \simeq \ell^* \frac{v}{r},
\]
whence $r(t) \simeq \sqrt{2\ell^* v t},$ and 
\[ \dot \theta \simeq \frac{v}{r} \simeq \sqrt{\frac{v}{2\ell^* t}}\]
whence $\theta(t) \simeq \sqrt{\frac{2 v t}{\ell^*}}.$ Together, these expressions for $r(t)$ and $\theta(t)$ describe an Archimedean spiral with asymptotically constant speed. Adapting them to the cone easily yields Eq.~5 from the main text.

As an aside, Archimedean spirals are observed both in the present work and previous studies, so it is tempting to apply the same simplified low-information model (Eq.~4 in the main text) to the isotropic problem. However, the isotropic problem lacks a substantial region where $p_1\simeq 1,$ so the model's validity is dubious and we cannot determine the characteristic lengthscale of the spiral with such an argument.

\subsection{A family of fast-decaying distributions which lead to subdiffusive scaling} 
It is also interesting to try to understand when Eq.~\ref{eq:constraint} is \emph{not} asymptotically saturated. Intuitively, if $p(r)$ decays sufficiently quickly, the extra time it takes to sweep out circles with larger radii will outweigh the benefit of sweeping out more probability mass per unit time, so the agent will be incentivized to move outward very slowly.

Again, we do not attempt to answer this question in full generality. However, in particular, one can show that compressed exponentials of the form $p(r) \propto \exp(-r^\beta)$ for $\beta>1$ have optimal $\tilde r_{k+1} - \tilde r_k$ tending toward 0. We suppress details since the result is a straightforward generalization of Lemmas 4.1--4.3 from Ref.~\cite{beck1984}, which studied the special case of a Gaussian distribution $p(r) \propto \exp(-r^2/2).$

The upshot is that if $p(r)$ decays faster than exponential, Eq.~\ref{eq:constraint} will \emph{not} be asymptotically saturated by an optimal policy, at least for one family of distributions which includes Gaussians. This, in turn, will lead to scaling in time slower than $r\sim t^{1/2}.$ (N.B., if $p(r)$ decays too quickly, then we risk breaking the approximation that allowed us to consider a prior supported only on a cone.)

\begin{figure}
    \centering
    \includegraphics[width=0.7\linewidth]{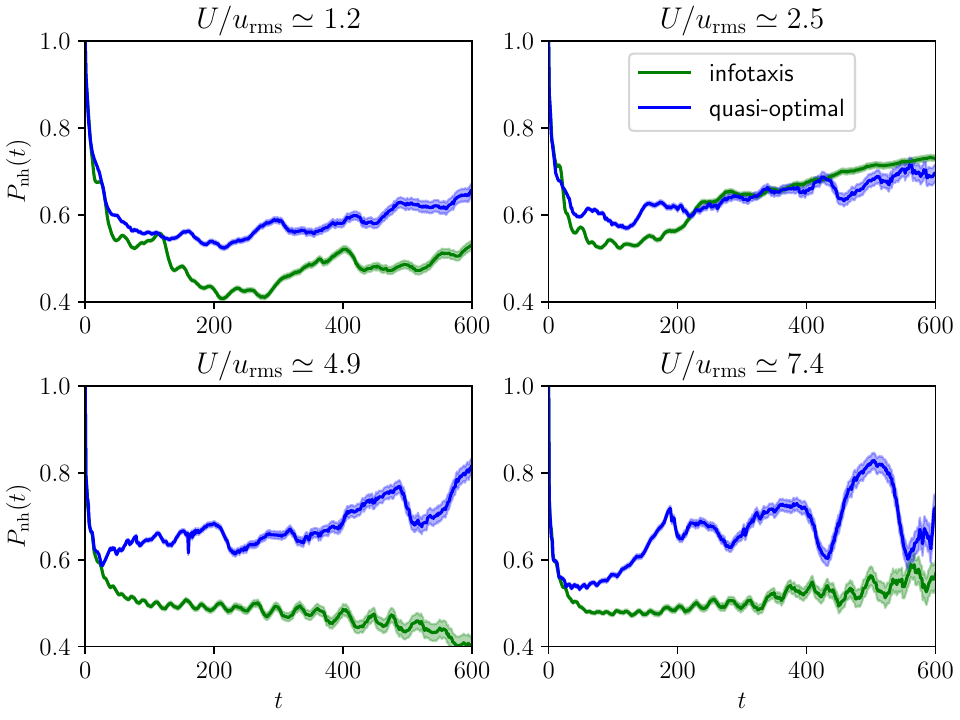}
    \caption{\RH{Probability $P_{\rm nh}(t)$ that, at time $t,$ the searcher has not encountered the cue since time 0, given that it has not yet found the source, estimated using an ensemble of $10^4$ trajectories and shown for each case with mean wind $U\ne0$. Error bars are shown as the standard error for a Bernoulli distribution.}}
    \label{fig:windy_no_hit_prob}
\end{figure}

\begin{figure}
    \centering
    \includegraphics[width=0.5\linewidth]{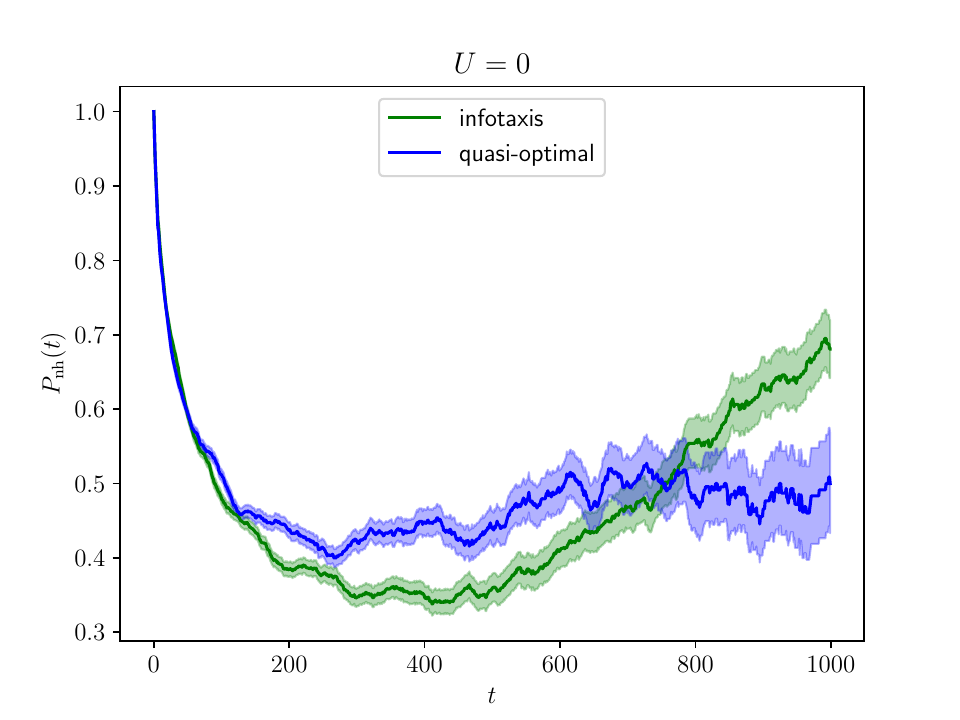}
    \caption{\RH{Same as Fig.~\ref{fig:windy_no_hit_prob}, but shown instead for the isotropic case $U=0$.}}
    \label{fig:iso_hit_prob}
\end{figure}

\begin{figure}
    \centering
    \includegraphics[width=\linewidth]{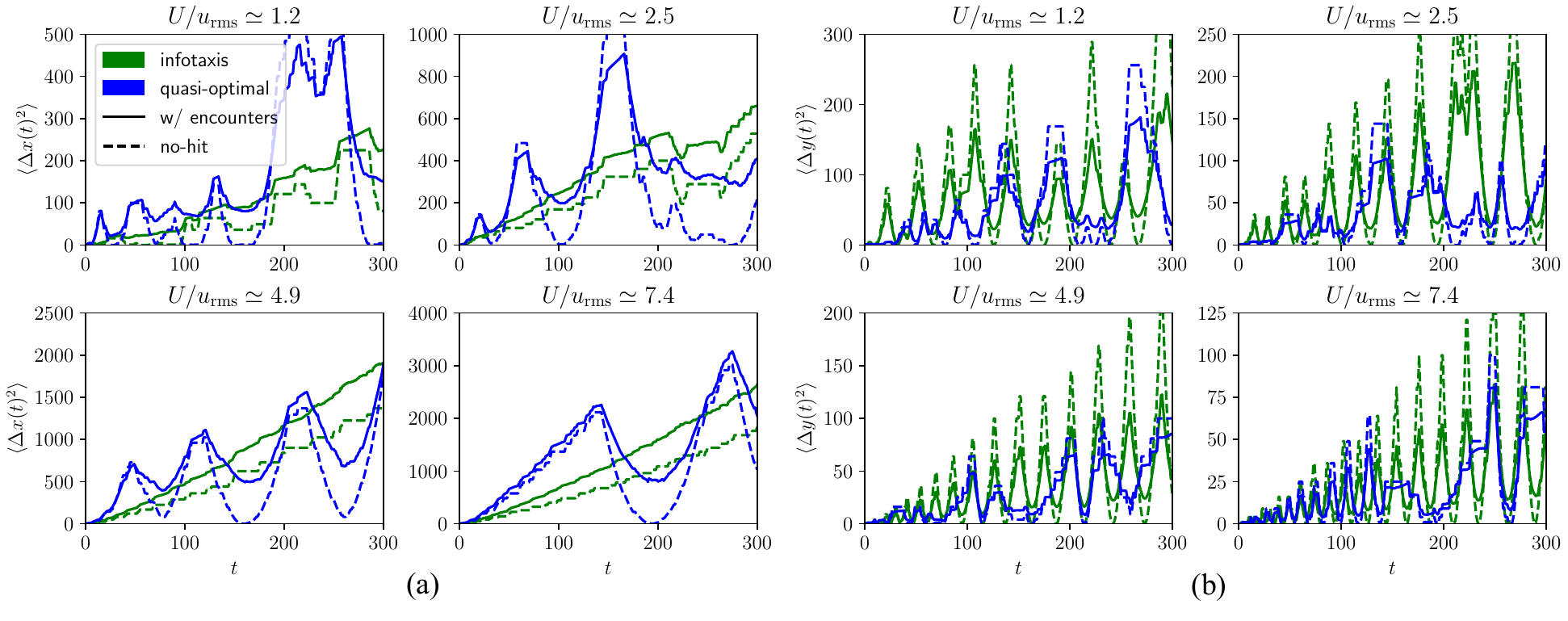}
    \caption{Mean squared displacements in $\hat{x}$ and $\hat{y}$ directions, for all nonzero wind speeds, taken from ensembles of $10^4$ trajectories and compared to the squared displacement in the no-hit trajectory (dashed lines). }
    \label{fig:windy_msd}
\end{figure}

\begin{figure}
\centering
\begin{subfigure}{.5\textwidth}
  \centering
  \includegraphics[width=\linewidth]{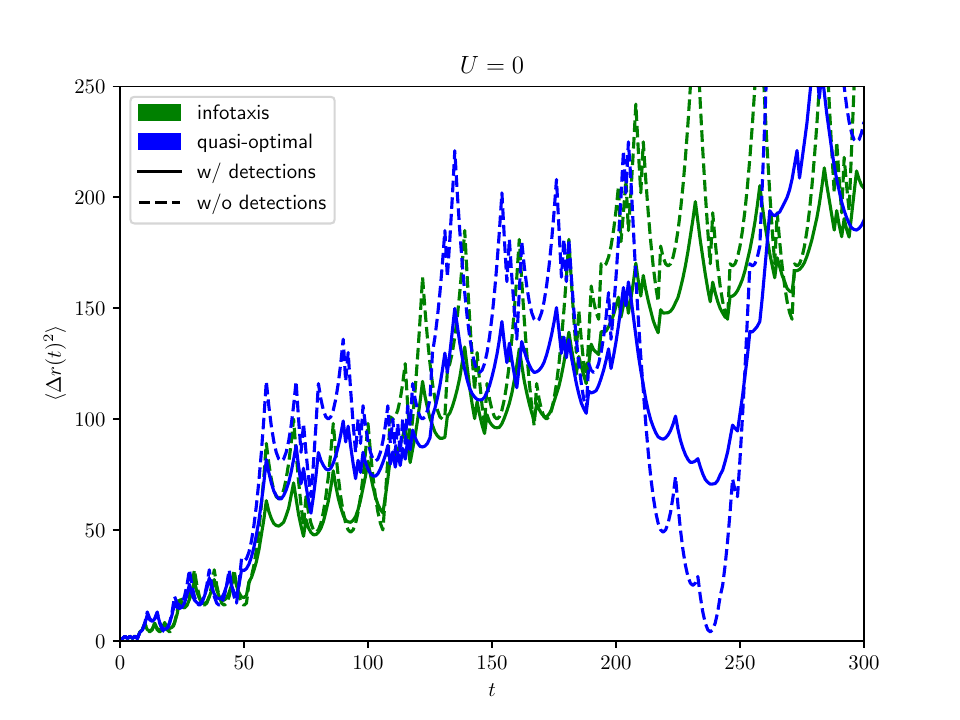}
  \caption{}
  \label{fig:sub1}
\end{subfigure}%
\begin{subfigure}{.5\textwidth}
  \centering
  \includegraphics[width=\linewidth]{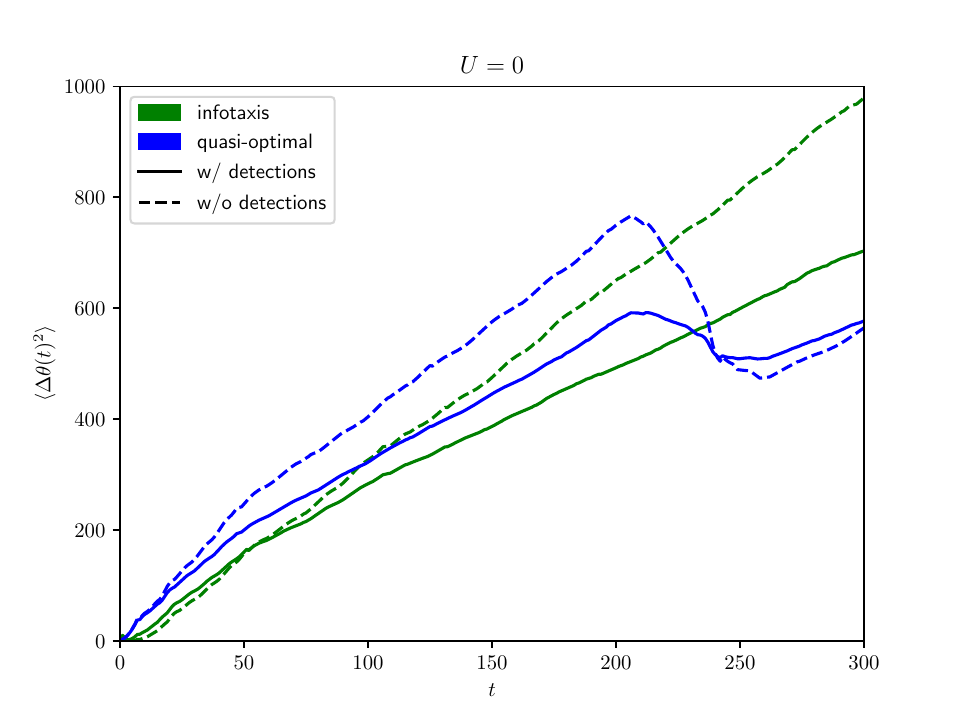}
  \caption{}
  \label{fig:sub2}
\end{subfigure}
\caption{(a) The mean squared displacement $\langle r(t)^2 \rangle = \langle x(t)^2 + y(t)^2 \rangle$ and  (b) the mean squared polar angle $\langle \theta(t)^2$ in the isotropic problem, taken from an ensemble of $10^4$ trajectories. The results are roughly consistent with $r\propto t^{1/2}$ and $\theta \propto t^{1/2},$ which describes an Archimedean spiral.}
\label{fig:isotropic_msd}
\end{figure}

\RH{\section{Additional plots and movies}}
\RH{\subsection{No-hit probability and MSD at all wind speeds}}
$P_{\rm nh}(t),$ as defined in the main text, is shown for all wind speeds in Figs.~\ref{fig:windy_no_hit_prob}--\ref{fig:iso_hit_prob}.  The mean squared displacement (MSD) data are presented in Figs.~\ref{fig:windy_msd}--\ref{fig:isotropic_msd}. In the isotropic case, we also show the mean squared polar angle $\langle \theta(t) \rangle$. As a reminder, an ensemble of $10^4$ trajectories for each wind speed was used to estimate the \RH{MSD, and an ensemble of size $10^5$ was used to estimated $P_{\rm nh}$ (except for the isotropic case, which only used $10^4$ due to the relatively long search time required).}

\begin{figure}
    \centering
    \includegraphics[width=0.7\linewidth]{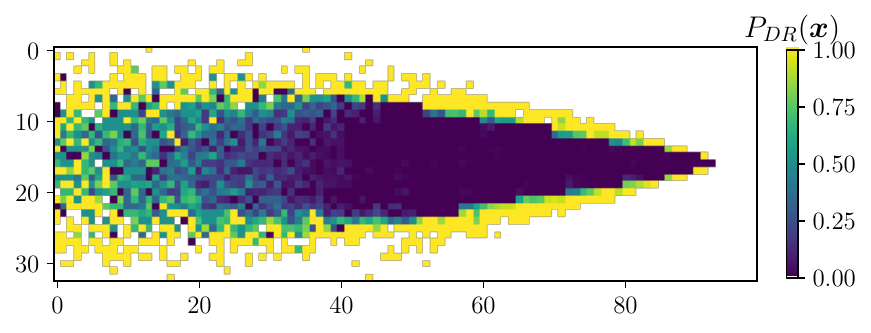}
    \caption{\RH{Probability $P_{\rm DR}(\bm x)$ that a quasi-optimal agent will eventually initiate a downwind return, given that it started its search at $\bm{x},$ when $U/u_{\rm rms} \simeq 7.4,$ computed from an ensemble of $10^5$ trajectories. This is defined as the probability that the agent goes $>150$ timesteps without an encounter after time zero.}}
    \label{fig:pdr}
\end{figure}

\RH{\subsection{Probability of executing a downwind return}
When $U/u_{\rm rms} \simeq 7.4,$ how often does a quasi-optimal agent actually execute a downwind return in practice? To answer this, in Fig.~\ref{fig:pdr} we plot the probability $P_{\rm DR}(\bm{x})$ that the agent goes $t_R=150$ timesteps without registering an encounter, as a function of the starting position --- and therefore at least begins a downwind return. In general, a downwind return is likely if the agent starts far off-axis (near the edge of the likelihood cone), or if the agent starts far downwind.}

\begin{figure}
    \centering
    \includegraphics[width=0.7\linewidth]{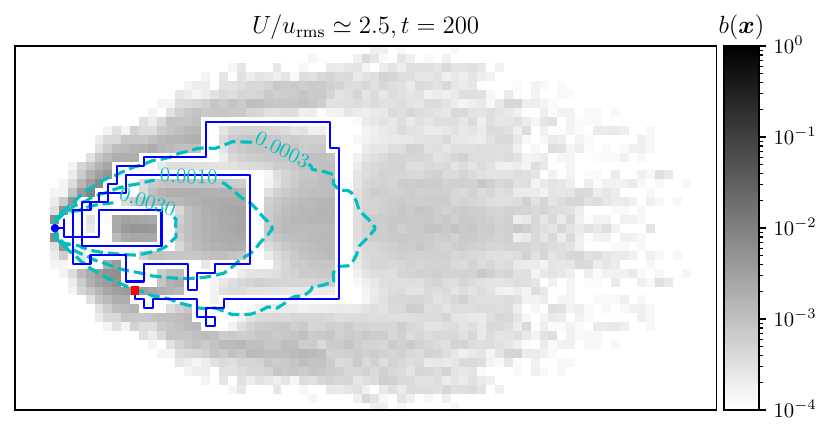}
    \caption{\RH{No-hit trajectory for $U/u_{\rm rms}\simeq 2.5,$ overlaid with three isolines of the prior.}}
    \label{fig:isolines}
\end{figure}

\begin{figure}
    \centering
    \includegraphics[width=0.5\linewidth]{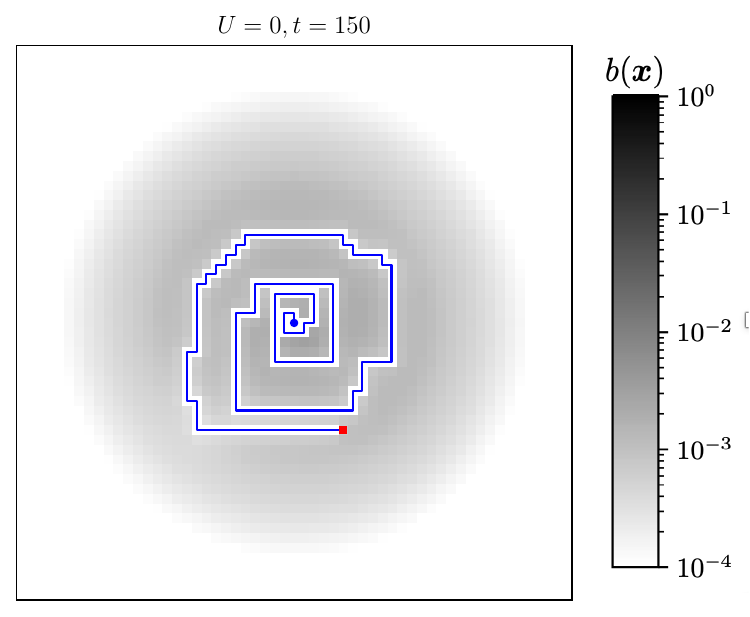}
    \caption{\RH{Quasi-optimal no-hit trajectory for the isotropic case $U=0.$}}
    \label{fig:v0_no_hit}
\end{figure}

\RH{\subsection{Additional no-hit trajectory plots}
In the main text, it was claimed that the quasi-optimal no-hit trajectory for $U/u_{\rm rms}\simeq 2.5$ appears to roughly follow isolines of the prior. We visualize this in Fig.~\ref{fig:isolines}, by overlaying a few isolines with the no-hit trajectory. We also claimed that for $U=0,$ the no-hit trajectory roughly resembles an Archimedean spiral. This can be inferred from Fig.~\ref{fig:isotropic_msd}, in which both $r^2$ and $\theta^2$ scale very roughly as linear in time (again, $r$ and $\theta$ are the polar coordinates with the starting point as the origin). For the reader's aid, we also include a visualization of the trajectory itself, Fig.~\ref{fig:v0_no_hit}. It should be noted that perfect agreement with an Archimedean spiral is impossible, due to radial symmetry breaking by the square grid.
}

\begin{figure}
    \centering
    \includegraphics[width=0.7\linewidth]{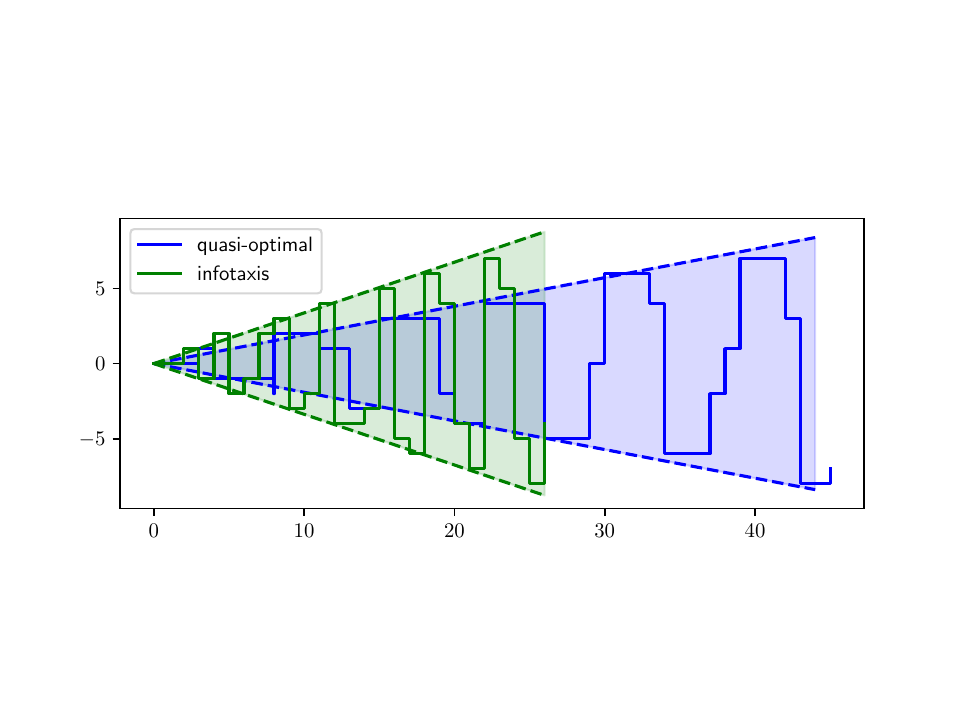}
    \caption{Quasi-optimal and infotactic no-hit trajectories for $U/u_{\rm rms} \simeq 7.4,$ along with a best fit to the cones of exploration, which have been outlined with dashed lines and shaded. The trajectories are shown at $t_R=150,$ the starting point for the quasi-optimal policy's downwind return.}
    \label{fig:cones}
\end{figure}

\RH{Finally, in Fig.~\ref{fig:cones} we overlay the trajectories quasi-optimal and infotactic no-hit trajectories for $U/u_{\rm rms} \simeq 7.4$. The trajectories are plotted up to the time $t_R=150,$ which marks the beginning of the downwind return in the quasi-optimal policy. We plot linear fits to the envelopes of the zigzag oscillations in order to visualize the cones of exploration of each policy. One can see clearly that the cone is significantly narrower for the quasi-optimal policy than for infotaxis, a signal that the former policy is greedier.} 

\RH{While the quasi-optimal policy statistically outperforms infotaxis by a substantial margin, in the unlikely event that the source lies within infotaxis' cone of exploration up to $t_R,$ but outside of the quasi-optimal policy's cone, it is probable that infotaxis will find the source first. 
}

\subsection{\RH{No-hit trajectory movies}}
Also included in the Supplemental Material as separate files are movies, in .gif format, of no-hit trajectories at all wind speeds, for both infotaxis and quasi-optimal trajectories. The beginning of the file name indicates the first digit of the normalized wind speed.